\documentclass[11pt]{article}

\usepackage{epsfig}

\RequirePackage{xspace}

\def\W      {\ensuremath{W}\xspace}

\def\Kbar  {\kern 0.2em\overline{\kern -0.2em K}{}\xspace}

\def\Kz    {\ensuremath{K^0}\xspace}
\def\Kzb   {\ensuremath{\Kbar^0}\xspace}
\def\KzKzb {\ensuremath{\Kz \kern -0.16em \Kzb}\xspace}
\def\Kp    {\ensuremath{K^+}\xspace}
\def\Km    {\ensuremath{K^-}\xspace}

\def\KpKm  {\ensuremath{\Kp \kern -0.16em \Km}\xspace}

\def\Dbar    {\kern 0.2em\overline{\kern -0.2em D}{}\xspace}

\def\Dz      {\ensuremath{D^0}\xspace}
\def\Dzb     {\ensuremath{\Dbar^0}\xspace}
\def\DzDzb   {\ensuremath{\Dz {\kern -0.16em \Dzb}}\xspace}
\def\Dp      {\ensuremath{D^+}\xspace}
\def\Dm      {\ensuremath{D^-}\xspace}

\def\DpDm    {\ensuremath{\Dp {\kern -0.16em \Dm}}\xspace}

\def\Dstarz  {\ensuremath{D^{*0}}\xspace}
\def\Dstarzb {\ensuremath{\Dbar^{*0}}\xspace}

\def\Bbar    {\kern 0.18em\overline{\kern -0.18em B}{}\xspace}

\def\Bz      {\ensuremath{B^0}\xspace}
\def\Bzb     {\ensuremath{\Bbar^0}\xspace}
\def\BzBzb   {\ensuremath{\Bz {\kern -0.16em \Bzb}}\xspace}
\def\Bu      {\ensuremath{B^+}\xspace}
\def\Bub     {\ensuremath{B^-}\xspace}
\def\Bp      {\ensuremath{\Bu}\xspace}

\def\BpBm    {\ensuremath{\Bu {\kern -0.16em \Bub}}\xspace}

\def\BorBbar    {\kern 0.18em\optbar{\kern -0.18em B}{}\xspace}
\def\DorDbar    {\kern 0.18em\optbar{\kern -0.18em D}{}\xspace}
\def\KorKbar    {\kern 0.18em\optbar{\kern -0.18em K}{}\xspace}

\mathchardef\Upsilon="7107
\def\Y#1S{\ensuremath{\Upsilon{(#1S)}}\xspace}

\mathchardef\Deltares="7101
\mathchardef\Xi="7104
\mathchardef\Lambda="7103
\mathchardef\Sigma="7106
\mathchardef\Omega="710A

\def\Deltabar{\kern 0.25em\overline{\kern -0.25em \Deltares}{}\xspace}
\def\Lbar{\kern 0.2em\overline{\kern -0.2em\Lambda\kern 0.05em}\kern-0.05em{}\xspace}
\def\Sigbar{\kern 0.2em\overline{\kern -0.2em \Sigma}{}\xspace}
\def\Xibar{\kern 0.2em\overline{\kern -0.2em \Xi}{}\xspace}
\def\Obar{\kern 0.2em\overline{\kern -0.2em \Omega}{}\xspace}
\def\Nbar{\kern 0.2em\overline{\kern -0.2em N}{}\xspace}
\def\Xb{\kern 0.2em\overline{\kern -0.2em X}{}\xspace}

\newcommand{\tev}{\ensuremath{\mathrm{\,Te\kern -0.1em V}}\xspace}
\newcommand{\gev}{\ensuremath{\mathrm{\,Ge\kern -0.1em V}}\xspace}
\newcommand{\mev}{\ensuremath{\mathrm{\,Me\kern -0.1em V}}\xspace}
\newcommand{\kev}{\ensuremath{\mathrm{\,ke\kern -0.1em V}}\xspace}
\newcommand{\ev}{\ensuremath{\mathrm{\,e\kern -0.1em V}}\xspace}
\newcommand{\gevc}{\ensuremath{{\mathrm{\,Ge\kern -0.1em V\!/}c}}\xspace}
\newcommand{\mevc}{\ensuremath{{\mathrm{\,Me\kern -0.1em V\!/}c}}\xspace}
\newcommand{\gevcc}{\ensuremath{{\mathrm{\,Ge\kern -0.1em V\!/}c^2}}\xspace}
\newcommand{\mevcc}{\ensuremath{{\mathrm{\,Me\kern -0.1em V\!/}c^2}}\xspace}

\def\mus  {\ensuremath{\rm \,\mus}\xspace}

\def\mus        {\ensuremath{\,\mu{\rm s}}\xspace}    

\def\to                 {\ensuremath{\rightarrow}\xspace}

\def\pep2{PEP-II}

\def\gsim{{~\raise.15em\hbox{$>$}\kern-.85em
          \lower.35em\hbox{$\sim$}~}\xspace}
\def\lsim{{~\raise.15em\hbox{$<$}\kern-.85em
          \lower.35em\hbox{$\sim$}~}\xspace}

\xspace

\def\jetset74   {\mbox{\tt Jetset \hspace{-0.5em}7.\hspace{-0.2em}4}\xspace}

\newcommand{\BDDK}{\ensuremath{B\to \Dbar^{(*)} D^{(*)} K}}
\newcommand{\BDDsK}{\ensuremath{B\to \Dbar D^* K}}
\newcommand{\BDsDK}{\ensuremath{B\to \Dbar^* D K}}
\newcommand{\BDsDsK}{\ensuremath{B\to \Dbar^{*} D^{*} K}}
\newcommand{\BDDKspec}{B \rightarrow \Dbar D  K  }
\newcommand{\bccs}{b \rightarrow c \overline{c} s  }

\begin{document}

\title {\Large \bf \boldmath Isospin analysis of $\BDDK$ decays 
\footnote{DAPNIA-03-366,hep-ph/0401014}}
\setcounter{footnote}{2}
\author{
Marco Zito \footnote{zito@hep.saclay.cea.fr}\\
{\it \small DSM/Dapnia/SPP, CEA-Saclay, 91191 Gif/Yvette, France} }



\maketitle


\begin{abstract}
The peculiar isospin properties of the $\bccs$ current 
lead to a rich set of isospin relations for the $\BDDK$ decays
which are presented here. Recent
high quality experimental data on the complete set of these decays 
(22 measurements) are 
analysed in this context, the
isospin relations are tested and the results for the isospin 
amplitudes are discussed. Large values of the strong phases are
suggested by the data. 
The comparison between the measured and expected branching fractions yields
a new measurement of the ratio of branching fractions  
$ \frac{Br(\Upsilon(4S) \rightarrow B^+ B^-)} {Br(\Upsilon(4S) \rightarrow
\Bz \Bzb)}=0.86 \pm 0.13$.
We finally discuss the implications of our findings for the 
measurement of 
$\sin(2 \beta)$ and $\cos(2 \beta)$ 
using these decays. 
\end{abstract}

\section{Introduction}

Given the difficulties in computing in a reliable and model-independent
way the $B$ meson decay amplitudes to hadronic final states, isospin relations are
a very general and useful tool to establish relations between
various $B$ decay modes.
The peculiar isospin properties of the $\bccs$ current 
are known since a long time
\cite{ref:sanda} and they have already been 
been used in the context of $B$ meson decays~\cite{ref:cleo}.
 The possibility that a large fraction of $\bccs$  decays hadronize
as $\BDDK$ was first suggested in Ref.~\cite{ref:buchalla}
 in the context of the discrepancy between the 
measured $B$ semi-leptonic rate
and the theoretical prediction. 
The same article suggested the use of isospin relations for the study of these decays. An additional motivation for an in-depth study of 
these channels is the possibility, originally discussed in 
Ref.~\cite{ref:CPDDK_1,ref:CPDDK_2,ref:CPDDK_3}, to measure
$\sin(2 \beta)$ and $\cos(2 \beta)$ 
using these decays.
Indeed they proceed through the same quark current than the 
gold-plated mode $\Bz \to J/\Psi \Kz$ and are not CKM-suppressed
to the difference of the $\Bz \to \bar D^{(*)} D^{(*)}$ modes. 

This Letter presents the complete set of isospin relations for 
$\BDDK$ decays; they are compared to the measurements through a  fit 
of the experimental data which determines the isospin amplitudes. 
These decays have been the object of
recent experimental investigations~\cite{ref:otherDDK,ref:babarDDK}.
The last study by the BABAR Collaboration presents a complete 
set of measurements (22 branching fractions have been measured) with
good accuracy which is the experimental basis of this paper.
An additional experimental complication is due to the fact that the 
branching ratios $ Br(\Upsilon(4S) \rightarrow B^+ B^-)$ and 
$Br(\Upsilon(4S) \rightarrow
\Bz \Bzb)$, needed to compare the neutral to charged $B$ meson
decays measured at an $e+ e^-$ machine operating 
at the $\Upsilon(4S)$ resonance, is not well known. 
This issue is adressed in this Letter. 

The aim of this study is manyfold:
\begin{itemize}
\item verify the isospin relations using a new large set
of experimental results;
\item offer some insight into the $\BDDK$ decay mechanism 
from the inspection of the isospin
amplitudes;
\item present a new measurement of the ratio of branching fractions
$ \frac{Br(\Upsilon(4S) \rightarrow B^+ B^-)} {Br(\Upsilon(4S) \rightarrow
\Bz \Bzb)}$;
\item discuss the implications of our findings for the 
measurement of 
$\sin(2 \beta)$ and
$\cos(2 \beta)$ using these decays.
\end{itemize}

\section{Isospin relations for $\BDDK$ decays}

\begin{figure}[htb]
\begin{minipage}{8cm}
\begin{flushleft}
\includegraphics[width=7cm]{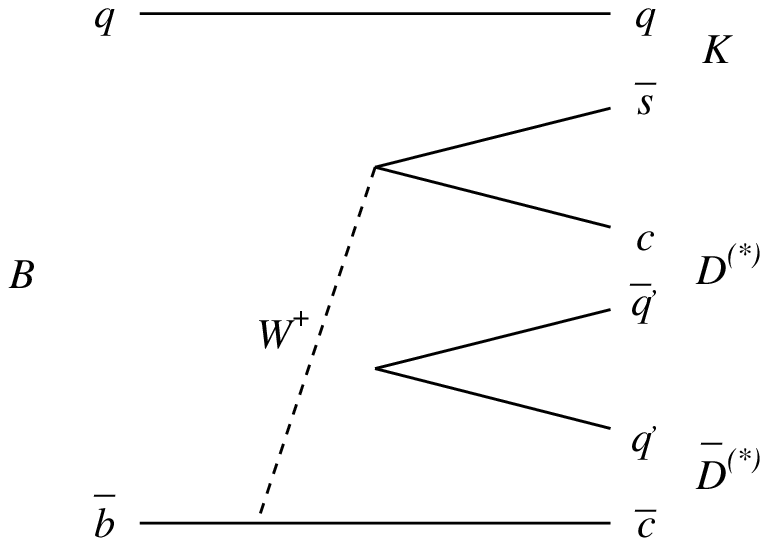}\\
\end{flushleft}
\label{fig:dstpi-diagrams-1}
\end{minipage}
\hfill
\begin{minipage}{8cm}
\begin{flushright}

\includegraphics[width=7cm]{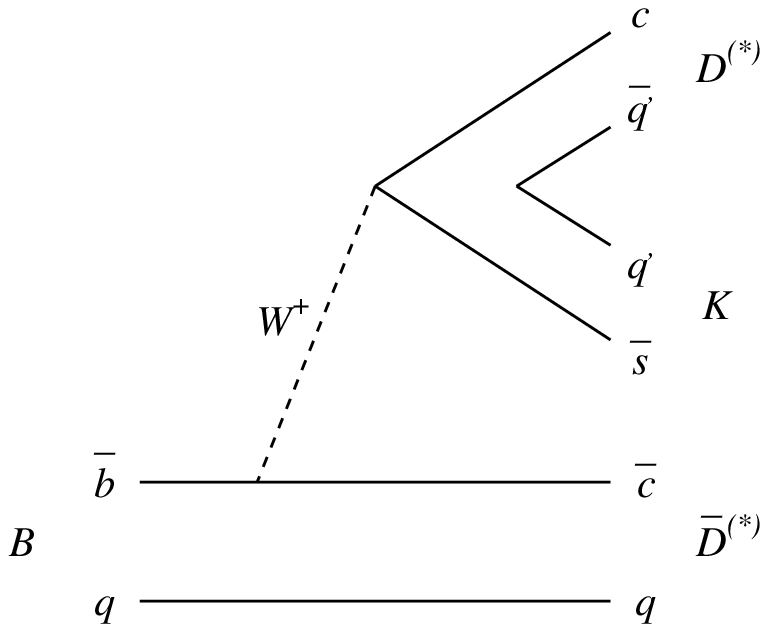}\\
\end{flushright}
\label{fig:dstpi-diagrams-2}
\end{minipage}
\caption{Left: internal \W-emission diagram for the decays
$B \to \Dbar^{(*)} D^{(*)} K$.
Right: external \W-emission diagram for the decays
$B \to \Dbar^{(*)} D^{(*)} K$.
  }
\label{Fi:diagrams}
\end{figure}

\begin{figure}[htb]
\begin{minipage}{8cm}
\begin{flushleft}
\includegraphics[width=7cm]{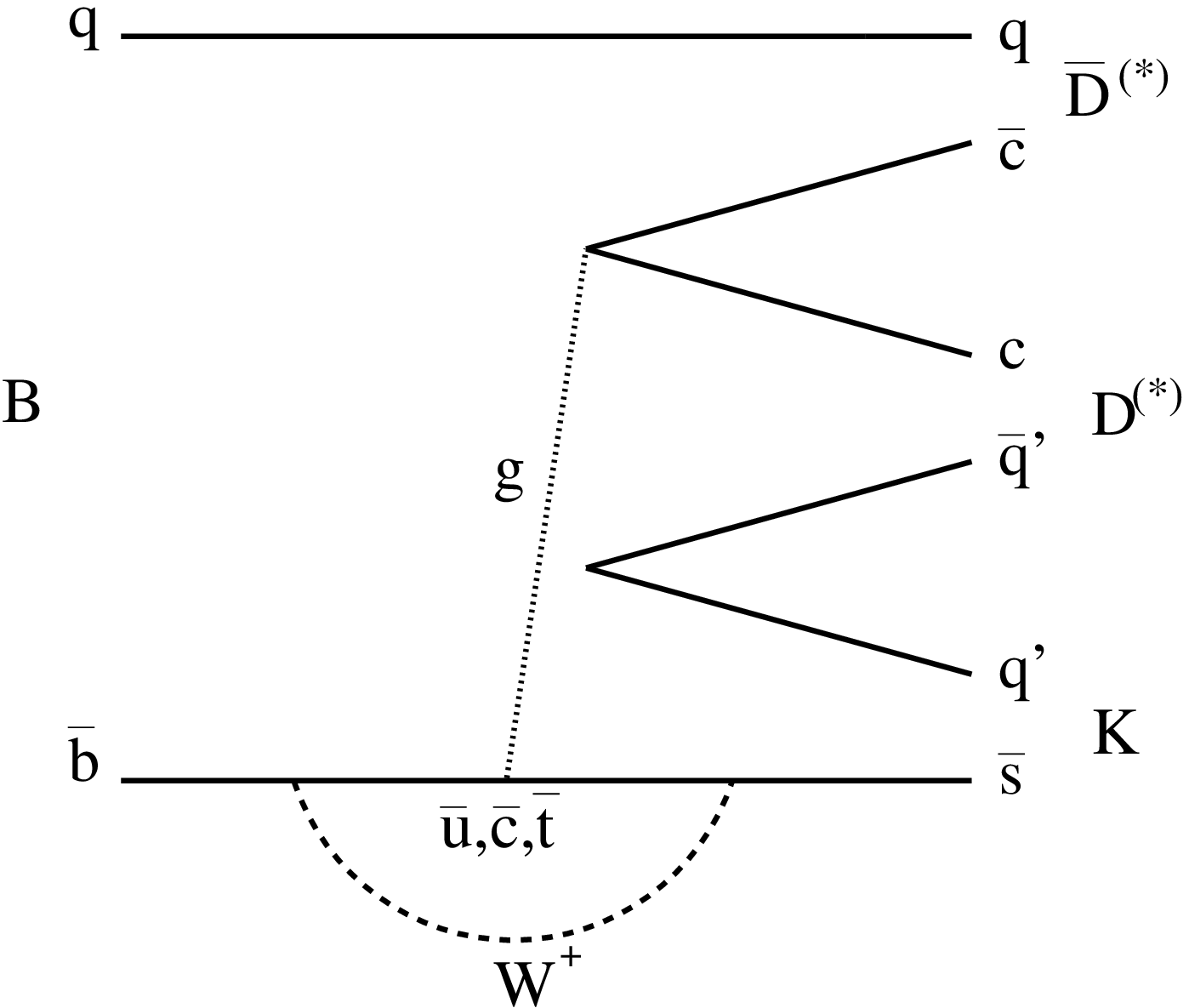}\\
\end{flushleft}
\end{minipage}
\hfill
\begin{minipage}{8cm}
\begin{flushright}
\includegraphics[width=7.cm]{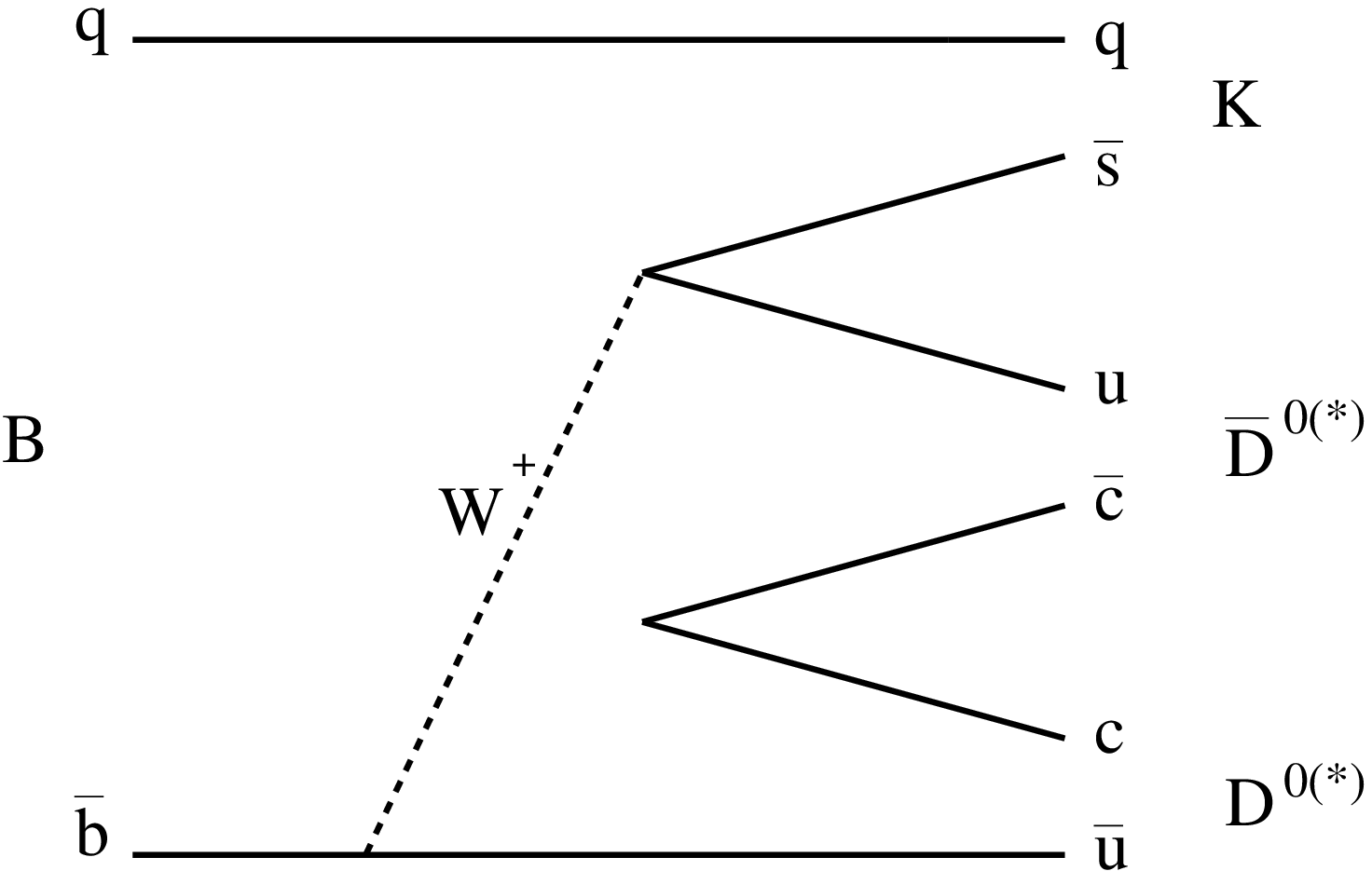}\\
\end{flushright}
\end{minipage}
\caption{Left: QCD penguin diagram  for the decays
$B \to \Dbar^{(*)} D^{(*)} K$.
Right: CKM suppressed diagram with $\Delta I = 1$ amplitude.
  }
\label{fig:supDiagram}
\end{figure}

The decays considered here are $\BDDK$, where $B$ is either a
$\Bz$ or $B^+$, and $K$ is either a $\Kz$ or $K^+$. These decays
proceed through a  $\bccs$ current through the diagrams of
Fig.~\ref{Fi:diagrams}. Depending on the final state, the external W-emission
diagram, the internal W-emission diagram (which is color-suppressed), or both
contribute to the transition amplitude. 
A penguin diagram, shown in Fig.~\ref{fig:supDiagram} (left plot),
can also contribute to the $\bccs$ current. It is expected to be
suppressed relative to the tree diagrams of Fig.~\ref{Fi:diagrams} 
and does not modify the isospin relations.

The decays $\Bz \to \Dbar^{(*)0} D^{(*)0} \Kz$ 
and $\Bp \to \Dbar^{(*)0} D^{(*)0} \Kp$
 could also proceed through a different diagram,
shown in Fig.~\ref{fig:supDiagram} (right plot), which could introduce a 
$\Delta I = 1$ amplitude. However this diagram proceeds through two
suppressed weak vertices $b \rightarrow u W$ and 
 $W \rightarrow s \bar u$ and a $c \bar c$ pair must be 
extracted from the vacuum, instead of a light quark pair as
in the CKM allowed diagrams. This amplitude is therefore 
suppressed by at least a factor $\lambda^2$, where 
$\lambda$ is the expansion parameter of the Wolfenstein parametrisation.
For these reasons we expect that $\Delta I = 0$
holds to  an excellent precision. 

As already mentioned, the isospin properties of the $\bccs$ current
are well known and follow from the fact that only isoscalar quarks
are involved. Therefore this is a $\Delta I = 0$ weak transition
and the final state is an isospin eigenstate.
The most general expression of these properties
 is given by the relation~\cite{ref:sanda} :
\begin{equation}
\Gamma(B^+ \rightarrow f(c \overline{c} s)) = \Gamma(\Bz \rightarrow
\tilde{f}(c \overline{c} s)),
\label{eq:basic}
\end{equation}
where $\tilde{f}(c \overline{c} s)$ is obtained from the state $f(c
\overline{c}
 s)$
through a $180^0$ isospin rotation.

While this relation applies to the $\BDDK$ decays, the full structure of
isospin relations can be obtained using the method described in
\cite{ref:peshkin}
and summarized here. Let us consider a N-particle state with individual
isospin
quantum numbers $I_k,m_k$ for $k=1,N$ ($N=3$ in our case). The isospin
wave function $ \psi(I,M)$ for a state of definite total isospin
(${\bf I} = \sum {\bf I_k}$) can be written as
 \begin{equation}
\psi(I,M) = \sum_t x_t(I)  \phi_t(I,M)
\end{equation}
where $t$ labels the invariant isospin quantum numbers 
$(t_2,t_3...,t_{N-1})$
of the operators ${\bf T}_k^2$ defined by
\begin{eqnarray}
{\bf T}_2 &=& {\bf I}_1 + {\bf I}_2, ... \nonumber \\
{\bf T}_k &=& {\bf T}_{k-1} + {\bf I}_k, ... \nonumber \\
{\bf I} &=&   {\bf T}_{N-1}+ {\bf I}_N
\end{eqnarray}
for $3 \le  k \le N-1$. The coefficients $ x_t(I)$ do not depend
on any isospin $z$ component and the basis functions $\phi_t(I,M) $
are simultaneous eigenfunctions of ${\bf I}^2, {\bf I}_z$ and all the 
${\bf T}^2_k$.
The amplitude for finding the state labeled by $m= (m_1,m_2...m_N)$
is
\begin{equation}
 <m | \psi(I,M) > = \sum_t x_t(I) U_{mt}(I,M)
\label{eq:amplitude}
\end{equation}
  where
\begin{eqnarray}
U_{mt}(I,M) & = & (I_1,m_1; I_2, m_2 | t_2, m_1+m_2) \nonumber \\
& \times &(t_2, m_1+m_2 ; I_3,m_3| t_3, m_1+m_2+m_3) \nonumber \\
&\times & ...(t_{N-1}, m_1+m_2...+m_{N-1} ; I_N,m_N| I, M ) 
\label{eq:clebsch}
\end{eqnarray}
and the terms on the right-hand side are Clebsch-Gordan coefficients.

In our case, just one operator ${\bf T}_2$ is introduced, 
with associated quantum numbers $t_2=0,1$. 
The equations \ref{eq:amplitude} and \ref{eq:clebsch}
generate the following set of relations 
\begin{eqnarray}
A(\Bz \rightarrow D^- \Dz K^+ ) &=& \frac{1} {\sqrt{6}}  A_1 -\frac{1} {\sqrt{2}}
A_0 \label{eq:isorel01}\\
A(\Bz \rightarrow D^- D^+ \Kz ) &=& \frac{1} {\sqrt{6}} A_1 +\frac{1} {\sqrt{2}}
A_0 \label{eq:isorel02}\\
A(\Bz \rightarrow \Dzb  \Dz \Kz )& =& - \sqrt{\frac{2} {3}}  A_1,
\label{eq:isorel03}
\end{eqnarray}
where $A_1$ ($A_0$) is the amplitude to produce the system $DK$
with isospin quantum number $t_2=1 (0)$. The $A_i$ amplitudes
in these formulae are equivalent to the $x_t(I)$ coefficients 
of Eq.~\ref{eq:amplitude}: they are reduced matrix elements, 
in the terms of the Wigner-Eckart theorem, of the isoscalar
Hamiltonian.

A similar set of relations holds for charged B meson decays
\begin{eqnarray}
A(B^+ \rightarrow \Dzb D^+ \Kz ) &=&  \frac{1} {\sqrt{6}}  A_1 -\frac{1} {\sqrt{2}}
A_0 \label{eq:isorelc1}\\
A(B^+ \rightarrow \Dzb \Dz K^+ ) &=& \frac{1} {\sqrt{6}} A_1 +\frac{1} {\sqrt{2}}
A_0 \label{eq:isorelc2}\\
A(B^+ \rightarrow D^-  D^+ K^+ )& =& - \sqrt{\frac{2} {3}}  A_1, 
\label{eq:isorelc3}
\end{eqnarray}
where the $A$ amplitudes are the same as for the neutral B decays.
This isospin decomposition of the $\BDDK$ amplitudes (Eq.~\ref{eq:isorel01} to~\ref{eq:isorelc3}) has already been
presented in Ref.~\cite{ref:pirjol} 
 where it has been discussed in the context of tests of factorization.
Identical equations hold for the other set of decays, \BDDsK, \BDsDK~and  \BDsDsK,
with different amplitudes $A$ in each case.  In the following we have used the superscripts $LL$, $L*$, $*L$ and $**$ 
for the $\BDDKspec$, $\BDDsK$, $\BDsDK$ and  $\BDsDsK$ decays
respectively.
Equivalent relations can be obtained 
considering the isospin quantum numbers of different subsytem of the 
final state ($D \overline{D}$, $\overline{D} K$). The $DK$ subsytem has been 
chosen here because in this case the transitions of Equations   
\ref{eq:isorel03} and \ref{eq:isorelc3},  proceeding only
through the color-suppressed 
diagrams of Fig.~\ref{Fi:diagrams} (left plot),  are 
associated only to the $A_1$ amplitude.

\begin{figure}[htb]
\begin{minipage}{8cm}
\begin{flushleft}
\includegraphics[width=7cm]{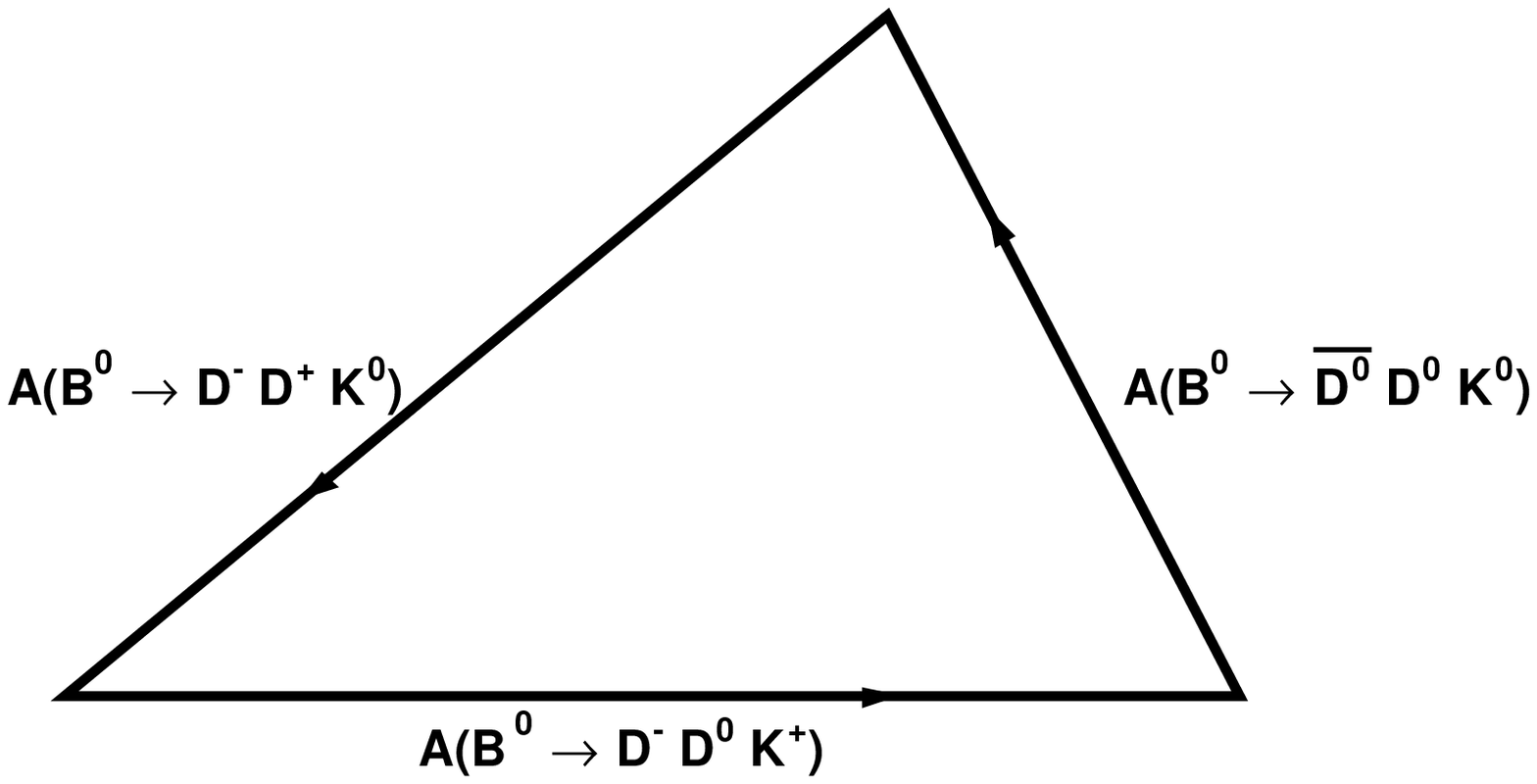}\\
\end{flushleft}
\end{minipage}
\hfill
\begin{minipage}{8cm}
\begin{flushright}
\includegraphics[width=7.cm]{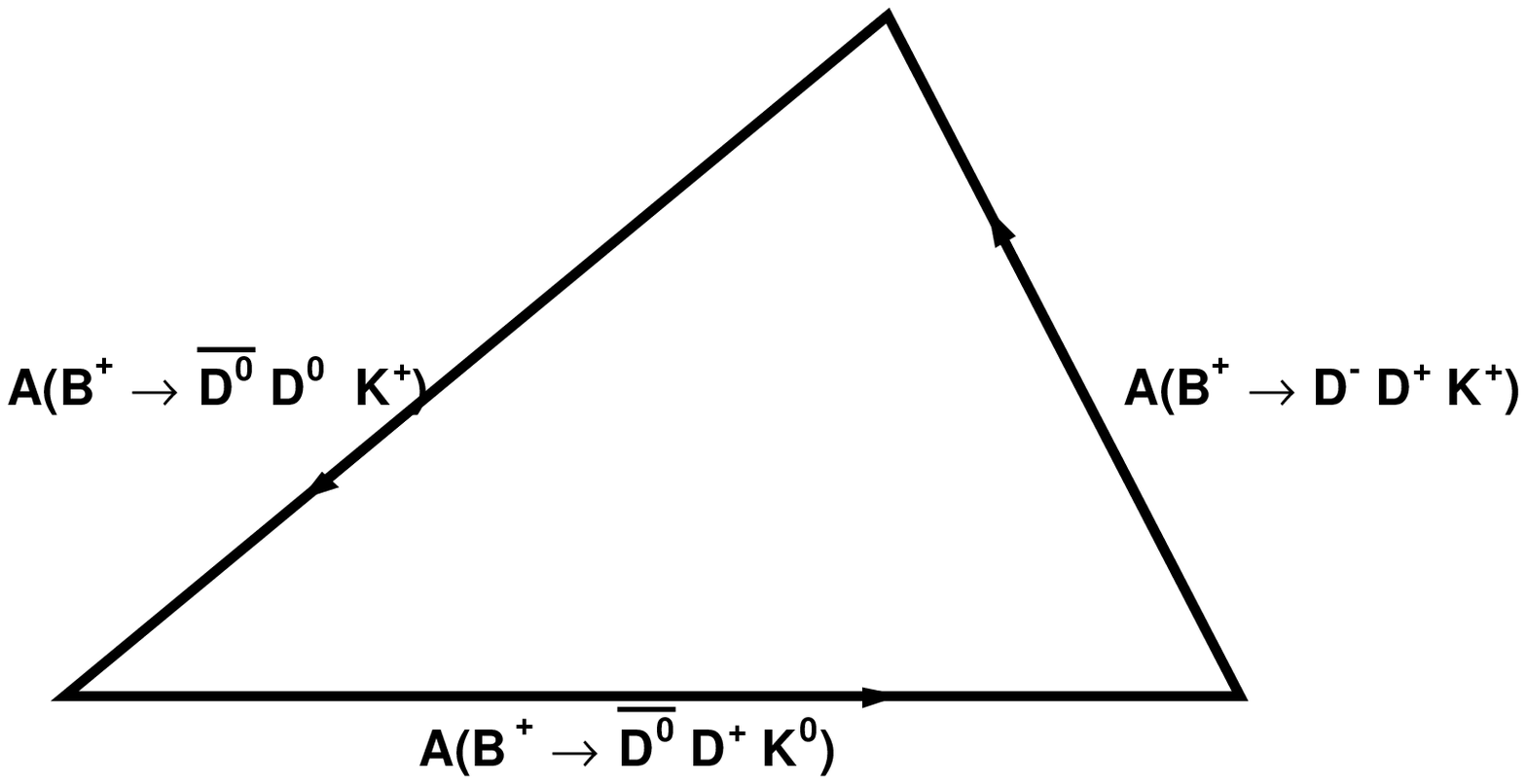}\\
\end{flushright}
\end{minipage}
\caption{Isospin triangles for the $\Bz$ (left) and $\Bp$ (right)
amplitudes.}

\label{fig:triangle_th}
\end{figure}

The relations presented above can be cast in the form of a triangle relation
between the amplitudes: 
\begin{eqnarray}
-A(\Bz \rightarrow D^- \Dz K^+ ) &=& A(\Bz \rightarrow D^- D^+ \Kz ) + A(\Bz \rightarrow \Dzb  \Dz \Kz ) \label{eq:triangle_bz}\\
-A(B^+ \rightarrow \Dzb D^+ \Kz ) &=& A(B^+ \rightarrow \Dzb \Dz K^+ ) + 
A(B^+ \rightarrow D^-  D^+ K^+ ) \label{eq:triangle_bp}
\end{eqnarray}
which are depicted in Fig.~\ref{fig:triangle_th}. 
The two triangles for $\Bz$ and $\Bp$ decays are identical according to the 
isospin relations, however experimentally it is advantageous to build the 
triangles separately with the $\Bz$ and $\Bp$ amplitudes. 

We finally notice that Eq.~\ref{eq:isorel01} to~\ref{eq:triangle_bp} 
are valid not only for the total decay amplitude but also for
each helicity amplitude separately as well as for  
the amplitude as a function of the Dalitz plot coordinates.

\section{Study of experimental results}

The branching fractions for the charged and neutral $B$ meson decay can be written 
\begin{eqnarray}
Br(\Bp \rightarrow f^+) &=& \tau_+  \frac{1}{(2 \pi)^3~32 M_B^3} \left ( \int dm^2_{D \bar D} 
dm^2_{D K} \right ) | A (\Bp \rightarrow f^+) |^2 \label{eq:branchingp}\\
Br(\Bz \rightarrow f^0) &=& \tau_0  \frac{1}{(2 \pi)^3~32 M_B^3} \left (\int dm^2_{D \bar D} 
dm^2_{D K}\right ) | A (\Bz \rightarrow f^0) |^2 ,
\label{eq:branching0}
\end{eqnarray}
where $\tau_+ = 2.543 \times 10^{12}~{\rm GeV^{-1}}$ and 
$\tau_0 = 2.343 \times 10^{12}~{\rm GeV^{-1}}$ \cite{ref:pdg}
are the lifetimes of the $\Bp$ and $\Bz$ mesons, 
$M_B$ is the mass of the B 
meson averaged over $\Bz$ and $\Bp$, $m_{D \bar D}$ and 
$m_{D K}$ are the invariant masses of the ${D \bar D}$ and 
${D K}$ subsystem,
 and the integral is computed numerically
over the allowed region of the three-body phase space. In computing these integrals 
the small mass differences between neutral and charged states for the $B$, $D^*$, $D$
and $K$ mesons have been neglected. 

The BABAR collaboration has recently studied the full set of $\BDDK$ decays
and has provided a measurement, reported in Table~\ref{tab:ddkyields}, for all these modes~\cite{ref:babarDDK}. 
These data, the  most precise to date, have been obtained at the PEP-II accelerator
from the reaction $e^+ e^- \rightarrow \Upsilon(4S) \rightarrow B \overline{B}$.
To compute the branching fractions it has been assumed that 
$Br(\Upsilon(4S) \rightarrow B^+ B^-) = Br(\Upsilon(4S) \rightarrow \Bz
\Bzb) = 0.5 $. However these equalities do not necessarily hold. 
 In order to account for
this factor, we have rewritten equations~\ref{eq:branchingp} and \ref{eq:branching0} 
in term of the rescaled amplitudes
$\tilde{A}  = \frac{A}{\sqrt{2 b_0}} $ 
where $b_0 =  Br(\Upsilon(4S) \rightarrow \Bz
\Bzb) $. The expression for $Br(\Bp \rightarrow f^+)$ is then multiplied
by the additional factor $f_{+/0}= \frac{Br(\Upsilon(4S) \rightarrow B^+
B^-)}{Br(\Upsilon(4S) \rightarrow \Bz \Bzb )}$. 

\begin{figure}[htb]
\begin{center}
        \includegraphics[width=15cm]{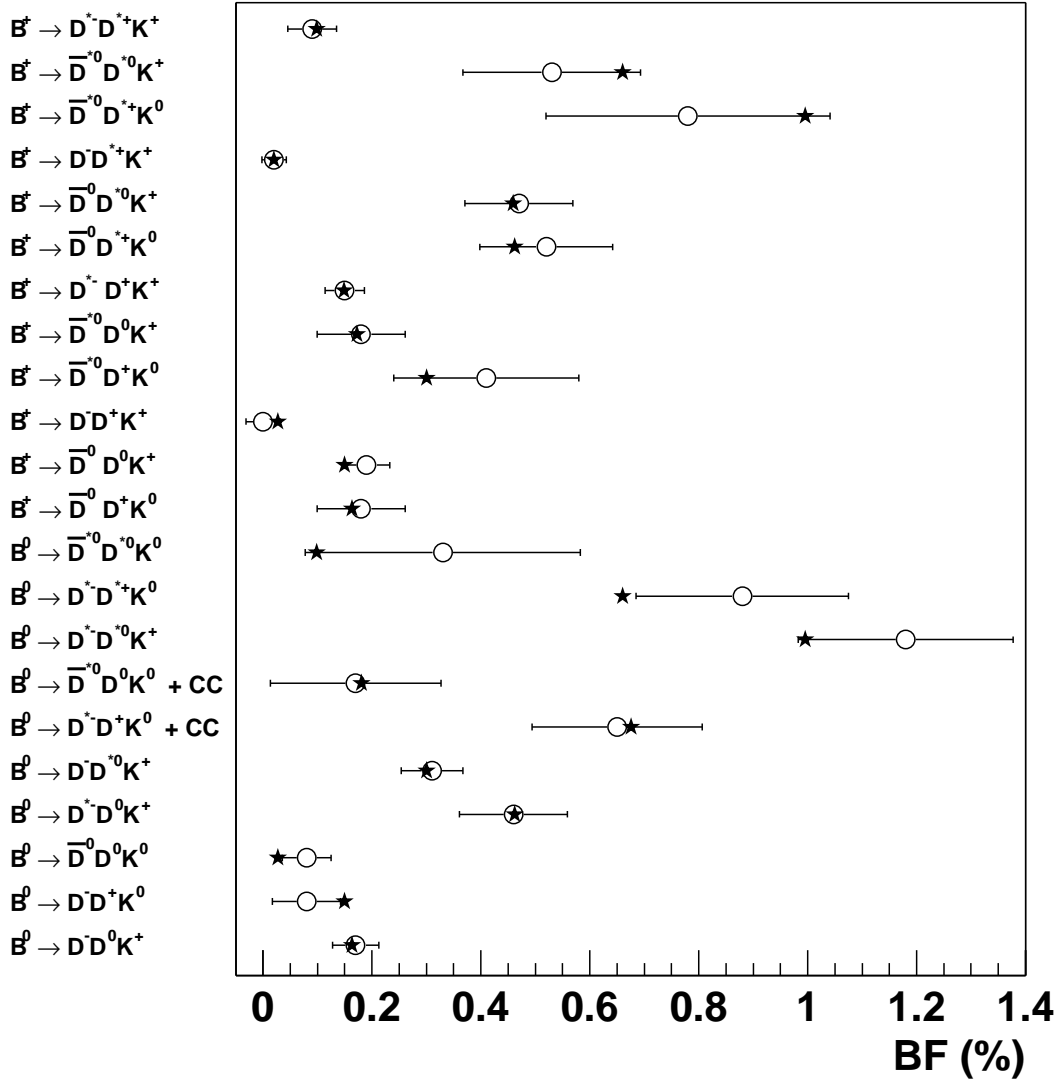}\\
\end{center}
\caption{Results of the $\chi^2$ fit to the experimental 
branching fractions. The fitted branching fractions 
are shown by the stars while the points with error bars show the
measured values.}
\label{fig:bfplot}
\end{figure}

The experimental data have been fitted simultaneously using the $\chi^2$
method where the 
fitted parameters are $f_{+/0}$ and  for each set of decays  $  |\tilde{A_1}|$, $
|\tilde{A_0}|$ 
and $ \delta = arg ( \tilde{A_1}
\tilde{A_0^*} )$. The total number of fitted parameters is 13. 
The results of the fit are reported in Tables~\ref{tab:ddkyields} 
and~\ref{table:fitres}. The
overall 
agreement between the measured and predicted branching fractions is good 
as can be judged from Table~\ref{tab:ddkyields}, Fig.~\ref{fig:bfplot} and
from 
the value $\chi^2 = 8.8 $ for 9 degrees of freedom ($n_{dof}$).
For this fit the statistical and systematical errors from 
Ref.~\cite{ref:babarDDK}
have been combined quadratically. This neglects the correlation between the systematical 
errors (common efficiencies, submode branching fractions, etc.).
For some $\Bz$ decays only the sum of the branching fraction with the
charge conjugate final state has been measured. We present
in Table~\ref{table:BFunsummed} the fitted values for the individual
branching fractions.  

\begin{figure}[htb]
\begin{center}
        \includegraphics[width=15cm]{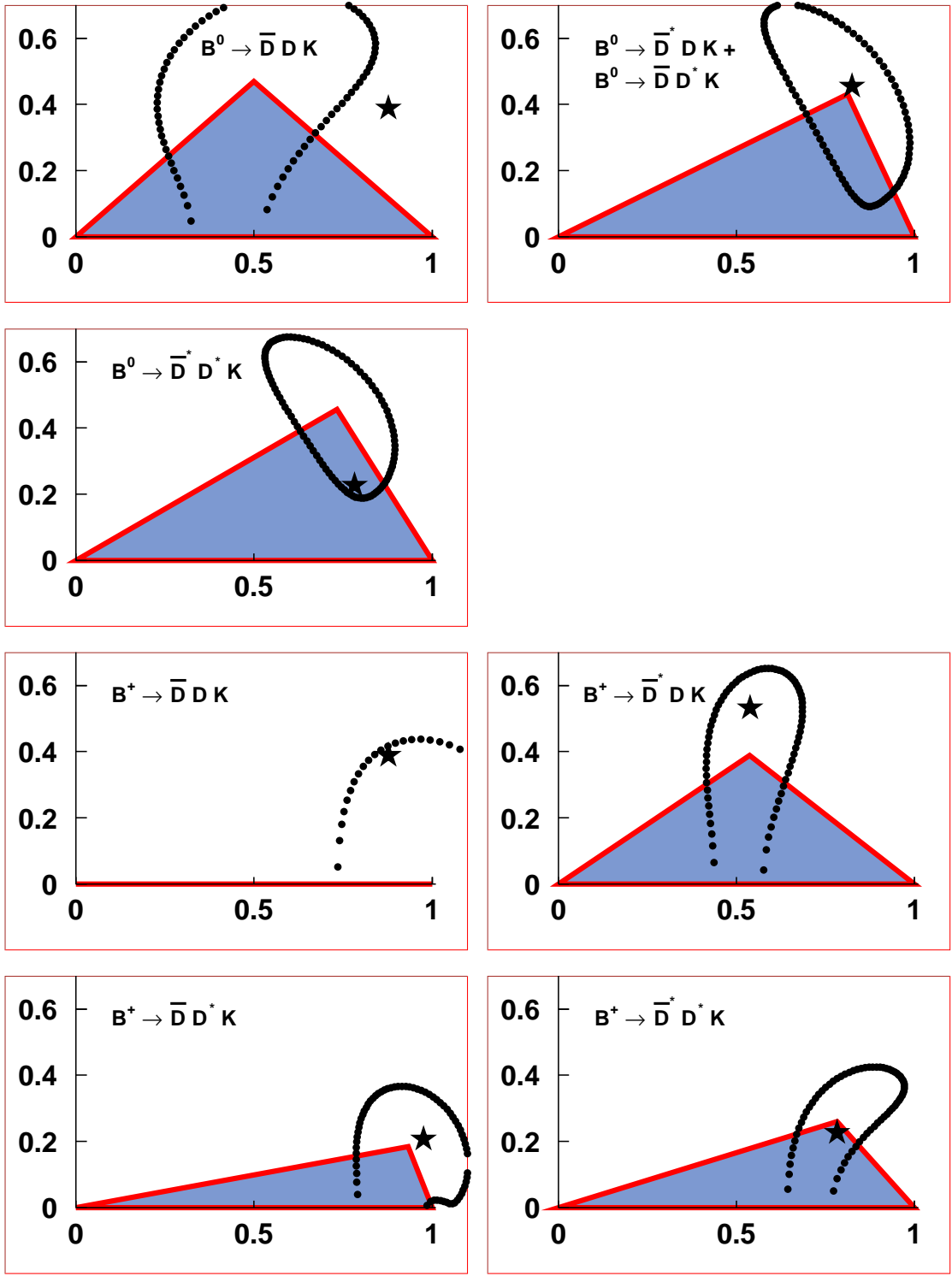}\\
\end{center}
\caption{Isospin triangles for the $\BDDK$ amplitudes. 
Each panel presents the measured vertex of the triangle, where the 
basis has been normalized to unity. The dotted contour shows the 
one standard deviation region. The star shows the result of the fit.  
We notice that only one triangle degenerates into a segment while
in all the other cases the shape of the triangle presents large angles.}
\label{fig:triangles_exp}
\end{figure}

An alternative way of displaying the experimental results and the fit results
is given by the isospin triangles introduced above.  
For ease of comparison, we have normalized the triangles to the size of 
the basis
($|A(\Bz \rightarrow D^{(*)-} D^{(*)0} K^+ )|$ and 
$|A(B^+ \rightarrow D^{(*)0}  D^{(*)+}  \Kz ) |$): 
therefore the lower side extends in each case from (0,0) to (1,0) and the
shapes of the triangles can be directly compared. Given that we have only
a measurement of the sides, there is a fourfold ambiguity on the vertex of 
the triangle. We have consistently chosen the same solution for its orientation. 
The seven measured triangles defined in this way are shown in 
Fig.~\ref{fig:triangles_exp} together with the fit result.

\begin{table*}[htb]
\caption{Branching fractions (BF) for each $\BDDK$ mode.
The first error on each branching fraction is the statistical uncertainty
and the second is the systematic uncertainty from \cite{ref:babarDDK}. 
The last column presents 
the result of the $\chi^2$ fit.}
\begin{center}
\begin{tabular}{|l|c|c|} \hline
  $B$ decay mode   & BF exp. $(\%)$ & BF fit $(\%)$   \\
\hline
\multicolumn{3}{|c|}{\Bz decays through external \W-emission amplitudes}
\\ \hline
 $\Bz\to \Dm \Dz\Kp$        & $0.17 \pm 0.03 \pm 0.03$ & 0.174 \\
$\Bz\to\Dm\Dstarz\Kp$       & $0.46 \pm 0.07 \pm 0.07$ &  0.495\\
$\Bz\to D^{*-} \Dz\Kp$      & $0.31^{+0.04}_{-0.03}\pm 0.04$ & 0.321\\
$\Bz\to D^{*-}\Dstarz\Kp$     & $1.18 \pm 0.10 \pm 0.17 $& 1.065 \\ \hline
\multicolumn{3}{|c|}{\Bz decays through external+internal
\W-emission amplitudes} \\ 
\hline
$\Bz\to\Dm\Dp\Kz$             & $0.08^{+0.06}_{-0.05}\pm 0.03$ & $0.161
$\\
$\Bz\to D^{*-}\Dp\Kz+\Dm D^{*+}\Kz$   & $0.65 \pm 0.12 \pm 0.10$ &  0.676
\\
$\Bz\to D^{*-} D^{*+}\Kz$     & $0.88^{+0.15}_{-0.14}\pm 0.13$ & 0.707 \\
\hline
\multicolumn{3}{|c|}{\Bz decays through internal \W-emission
amplitudes}
 \\ \hline
$\Bz\to \Dzb \Dz \Kz$       & $0.08 \pm 0.04 \pm 0.02$ & $0.029 $ \\
$\Bz\to\Dzb \Dstarz \Kz+ \Dstarzb \Dz \Kz$  & $0.17^{+0.14}_{-0.13}\pm
0.07$ & 0.181   \\
$\Bz\to \Dstarzb \Dstarz \Kz$  & $0.33^{+0.21}_{-0.20}\pm 0.14$ & $0.105 $ \\
\hline
\multicolumn{3}{|c|}{\Bu decays through external \W-emission amplitudes}
 \\ \hline
$\Bu\to \Dzb \Dp\Kz$         & $0.18 \pm 0.07 \pm 0.04$ & $0.163$ \\
$\Bu\to \Dstarzb\Dp\Kz$    & $0.41^{+0.15}_{-0.14}\pm 0.08$ & $0.300$ \\
$\Bu\to \Dzb D^{*+}\Kz$     & $0.52^{+0.10}_{-0.09}\pm 0.07$ & 0.462 \\
$\Bu\to \Dstarzb D^{*+}\Kz$  & $0.78^{+0.23}_{-0.21}\pm 0.14$ & 0.995\\
\hline 
\multicolumn{3}{|c|}{\Bu decays through external+internal
\W-emission amplitudes} \\ 
\hline
$\Bu\to \Dzb \Dz \Kp$       & $0.19 \pm 0.03 \pm 0.03$       & 0.150 \\
$\Bu\to\Dstarzb\Dz\Kp$   & $0.18^{+0.07}_{-0.06} \pm 0.04$ & $0.172$ \\
$\Bu\to \Dzb \Dstarz\Kp$   & $0.47 \pm 0.07 \pm 0.07$       & 0.459 \\
$\Bu\to\Dstarzb\Dstarz\Kp$   & $0.53^{+0.11}_{-0.10} \pm 0.12$ &0.660 \\
\hline
\multicolumn{3}{|c|}{\Bu decays through internal \W-emission amplitudes}
 \\ \hline
$\Bu\to \Dm\Dp\Kp$ & $0.00 \pm 0.03 \pm 0.01$ & $0.027 $ \\
$\Bu\to\Dm D^{*+}\Kp$  & $0.02 \pm 0.02 \pm 0.01$ & $0.020$ \\
$\Bu\to D^{*-} \Dp\Kp$  & $0.15 \pm 0.03 \pm 0.02$ & 0.149      \\
$\Bu\to D^{*-} D^{*+}\Kp$  & $0.09 \pm 0.04 \pm 0.02$ & $0.098$ \\ \hline
\end{tabular}
\end{center}
\label{tab:ddkyields}
\end{table*}


\begin{table*}[tb]
\caption{Results of the $\chi^2$ fit to the experimental 
branching fractions. The superscripts $LL$, $L*$, $*L$ and $**$ 
are for the $\BDDKspec$, $\BDDsK$, $\BDsDK$ and  $\BDsDsK$ decays
respectively. The amplitude values are in  units of $10^{-5}$ while the
phases $\delta$ are in degrees. The last column presents the results of the
fit introducing a constraint related to other measurements of 
$f_{+/0}$.
}
\begin{center}
\begin{tabular}{|l|c|c|} \hline
parameter & value & value \\
\hline
$|\tilde{A}_1^{LL}|$ & $ 0.28 \pm 0.13 $ & $ 0.25 \pm 0.13 $\\
$|\tilde{A}_0^{LL}|$ & $ 0.75 \pm 0.07 $ & $ 0.73 \pm 0.06 $\\ 
$\delta^{LL}$ & $ 95 \pm 22 $  & $ 100 \pm 23 $\\
\hline
$|\tilde{A}_1^{L*}|$ & $ 0.27 \pm 0.15 $ & $ 0.25 \pm 0.11 $\\
$|\tilde{A}_0^{L*}|$ & $ 1.51 \pm 0.11 $ & $ 1.45 \pm 0.09 $\\
$\delta^{L*}$ & $ 91 \pm 34 $ & $ 98 \pm 36 $\\
\hline
$|\tilde{A}_1^{*L}|$ & $ 0.75 \pm 0.10 $ & $ 0.69 \pm 0.08 $\\
$|\tilde{A}_0^{*L}|$ & $ 1.00 \pm 0.11 $ & $ 0.99 \pm 0.10 $\\
$\delta^{*L}$ & $ 111 \pm 17 $ & $ 116 \pm 14 $\\
\hline
$|\tilde{A}_1^{**}|$ & $ 0.71 \pm 0.17 $ & $ 0.66 \pm 0.14 $\\
$|\tilde{A}_0^{**}|$ & $ 2.38 \pm 0.17 $ & $ 2.27 \pm 0.14 $\\
$\delta^{**}$ & $ 127 \pm 26 $ & $ 133 \pm 22 $\\
\hline
$f_{+/0}$ & $ 0.86 \pm 0.13 $ & $ 1.02 \pm 0.05 $ \\
 \hline
$\chi^2/ n_{dof} $ & 8.8/9 & 10.4/10 \\
$Prob(\chi^2, n_{dof}) $ & 0.456 & 0.406  \\
 \hline
\end{tabular}
\end{center}
\label{table:fitres}
\end{table*}

\begin{table*}[htb]
\caption{ Fitted values of the branching fractions
for the $\BDDsK$ and $\BDsDK$ decays which have not been measured
individually.} 
\begin{center}
\begin{tabular}{|l|c|} \hline
$B$ decay mode & BF fit $(\%)$   \\ 
\hline
$\Bz\to D^{*-}\Dp\Kz $ & 0.185 \\ 
$\Bz\to \Dm D^{*+}\Kz$    &  0.491 \\
$\Bz\to \Dstarzb \Dz \Kz$ &  0.160 \\
$\Bz\to\Dzb \Dstarz \Kz$  &  0.021 \\
\hline
\end{tabular}
\end{center}
\label{table:BFunsummed}
\end{table*}

\clearpage

\section{Discussion}

\subsection{The value of $f_{+/0}$ and the validity of isospin relations}

The value of $f_{+/0}$ returned by the fit is
\begin{equation}
f_{+/0} = 0.86 \pm 0.13.
\label{eq:fpDDK}
\end{equation}
This value is in agreement with the theoretical predictions for $f_{+/0}$ which lie in the 
1.05-1.18 interval~\cite{ref:fptheor}, as well as 
with other determinations of this quantity: 
$f_{+/0} = 1.04 \pm 0.07 \pm 0.04$~\cite{ref:cleo}
and $f_{+/0} = 1.10 \pm 0.06 \pm 0.05$~\cite{ref:fpbabar} derived from similar
isospin relations for the branching fractions of $B$ decays to charmonium
final states.
Combining these measurements obtained in $B\rightarrow J/\Psi K$ decays,
 rescaled using the value 
$\tau_+/\tau_0 = 1.083 \pm 0.017 $~\cite{ref:pdg}, 
we obtain $
f_{+/0} = 1.046 \pm 0.056$.
We have added this constraint to the fit to the data obtaining 
the result shown in the last column of Table~\ref{table:fitres}. 
We notice that the measurement presented here does not improve substantially 
the uncertainty on $f_{+/0}$ and that the values and uncertainties 
on the amplitudes and phases do not change significantly using this constraint. 

The point can be investigated further. The inspection of Fig.~\ref{fig:bfplot} and 
Table~\ref{tab:ddkyields} shows that the branching fractions for \BDsDsK~decays
deviate from the fitted values in a correlated way. We have repeated the fit
separately for the three groups of decays final states obtaining the values for  
$f_{+/0}$ shown in Table~\ref{table:fp0sep}. We notice that the 
value measured in \BDsDsK~decays deviates from the experimental value 
in $B\rightarrow J/\Psi K$ decays
by 2.95 standard deviations. 

\begin{table*}[htb]
\caption{ Values of $f_{+/0}$ for the different groups of decay final states. 
 } 
\begin{center}
\begin{tabular}{|l|c|} \hline
final states & $f_{+/0}$ \\
\hline
$\BDDKspec$ & $1.24 \pm 0.43$  \\ 
$\BDsDK + \BDDsK$ & $1.01 \pm 0.21$  \\ 
$\BDsDsK$ & $0.55 \pm 0.16 $  \\ 
\hline
\end{tabular}
\end{center}
\label{table:fp0sep}
\end{table*}


This discrepancy can be explained either by 
an additional systematical effect in these measurements
or by a violation of the 
isospin symmetry for these final states. 
Clearly more data are needed to clarify this point. A high precision test 
of the isospin relations
will only be possible when  $f_{+/0}$ will be measured using a 
different experimental method. The large data sample accumulated
by the BABAR and BELLE experiments will allow this measurement 
in the near future. 


\subsection{Dynamical features of the amplitudes}

The amplitudes and phases extracted from the data present
some distintive features. 
First, within each set, the amplitude related to the color-suppressed decays
is much smaller, as expected. The ratios $A_0/A_1$ are presented
in Table \ref{table:aratio}. 
These ratios, except for the case of $\BDsDK$, are close to the 
na\"{\i}ve expectation $ |A_0|/|A_1|=N_c=3$, 
where $N_c$ is the number of colors. 

Second, the central values for the relative phases $\delta$ are 
in all cases close to $90^{\rm o}$. The errors on these values given in 
Table~\ref{table:fitres} are not relevant to determine confidence intervals 
because of the non-linear relation between $\delta$ and $\cos(\delta)$
which enters the $\chi^2$ expression. To do this the  $\chi^2$ profile
has been studied keeping in turn one phase $\delta$ fixed 
and repeating the fit.
The 90\% level confidence intervals are $ 92^{\rm o} < \delta^{*L} < 154^{\rm o}$ and 
$ 88^{\rm o} < \delta^{**} < 180^{\rm o}$ while no bound can be set for 
$ \delta^{LL}$ and $ \delta^{L*}$. The superscripts $LL$, $L*$, $*L$ and $**$ 
are for the $\BDDKspec$, $\BDDsK$, $\BDsDK$ and  $\BDsDsK$ decays
respectively. From this we can conclude that there is a 
reasonable indication for large strong phases in these amplitudes.
This suggests the presence of non-negligible Final State Interaction
for these decays. This is both an important indication {\it per se}
and has also consequences for the CP violation studies that will be
discussed in the next section. 

\begin{table*}[htb]
\caption{ Ratios $A_0/A_1$ from the fit to the data. } 
\begin{center}
\begin{tabular}{|c|c|} \hline
ratio & value \\
\hline
$ |A_0^{LL}|/|A_1^{LL}|$ & $2.68 \pm 2.44 $  \\
$ |A_0^{L*}|/|A_1^{L*}|$ & $5.59 \pm 2.04 $ \\
$ |A_0^{*L}|/|A_1^{*L}|$ & $1.33 \pm 0.24 $ \\
$ |A_0^{**}|/|A_1^{**}|$ & $3.35 \pm 0.98 $ \\
\hline
\end{tabular}
\end{center}
\label{table:aratio}
\end{table*}

\subsection{Implications for a $\sin(2\beta)$, $\cos(2\beta)$
measurement}

All the $\Bz \to \Dbar^{*} D^{*} \Kz$ are in principle good candidates
for the measurement of $\beta$. In the past the emphasis has been 
placed on the $\Bz \to \Dbar^{*} D^{*} \Kz$ and 
$\Bz \to \Dbar^{*} D^{*} \Kz$ decays 
\cite{ref:CPDDK_1,ref:CPDDK_2,ref:CPDDK_3} 
and preliminary theoretical values of the branching fractions have been presented.
We notice that the values for the branching fractions of these modes
presented in Tables~\ref{tab:ddkyields} and \ref{table:BFunsummed} 
can be used for more
precise assessments of the sensitivity of a measurement of 
$\beta$ using these modes. 

In  Ref.\cite{ref:babarDDK}, the observation of the modes
$\Bz \to D^{*-} D^{*+} \Kz$ and $\Bz \to D^{-} D^{*+}  \Kz +CC$
 is reported. We notice that for $\Bz \to D^{*-} D^{*+} \Kz$, the 
measured value of the branching fraction ($0.88^{+0.15}_{-0.14}\pm 0.13$)
and the value predicted by our fit ($0.707$) are almost a factor
two lower that what anticipated in Ref.~\cite{ref:CPDDK_3}, 
thereby  unfortunately also reducing the comparative advantage of this mode
with respect $\Bz \to D^{*-} D^{*+}$.

For $\Bz \to D^{-} D^{+} \Kz$, Ref.~\cite{ref:babarDDK} reports
only a 90\% CL upper limit (0.17 \%) which is very close to the fitted
value 0.161 \%. This means that the observation of this mode in the near future
is  possible. The estimated value of  Ref.~\cite{ref:CPDDK_2} 
($9~10^{-3}$) is a factor 6 above our predicted value.
We stress that this channel is a good candidate for CP-violation studies
because of the nature of the final state with three pseudoscalar particles. 
This will facilitate the angular analysis to determine the 
helicity amplitudes. 

Finally we stress that the $\Bz \to D^{*-} D^{+} \Kz$ and 
$\Bz \to D^{-} D^{*+} \Kz$ lead to final states accessible 
by both $\Bz$ and $\Bzb$. They can therefore be analysed in the same
way as described in Ref.\cite{ref:rhopi}. The strong phases play an 
important role for this analysis as the time-dependent
CP-asymmetry amplitudes are proportional to $\sin(2 \beta \pm \delta')$,
where $\delta'$ is the strong phase difference between 
$A(\Bz \to D^{-} D^{*+} \Kz)$ and $A(\Bzb \to D^{-} D^{*+} \Kz)$.
The possibly large values of the strong phases noticed above need to be taken
into account for any estimate of the sensitivities of this analysis.

\section{Conclusion}

We have presented the complete isospin relations for the $\BDDK$ decays. 
These relations have been compared to the recent experimental 
measurements through a fit of the isospin amplitudes. The overall agreement 
between the measured and the expected branching fractions is good
with the exception of a possible discrepancy for the \BDsDsK~decays. 
The isospin amplitudes present several peculiar features which 
point to a dynamical origin. Large values of the strong phases are
suggested by the data. 
We have also presented a new measurement of 
$ \frac{Br(\Upsilon(4S) \rightarrow B^+ B^-)} {Br(\Upsilon(4S) \rightarrow
\Bz \Bzb)} = 0.86 \pm 0.13$ 
in agreement with 
other determinations of this quantity. 
The implications of these results for the 
measurement of 
$\sin(2 \beta)$ and $\cos(2 \beta)$ 
using these decays have been discussed. 

\section{Acknowledgments}

The author wishes to warmly thank J. Charles, J.P. Lees and 
L. Oliver for reading the manuscript and making useful suggestions.

\end{document}